\documentclass[preprint]{aastex}

\slugcomment{\today }

\shorttitle{A New SX Phe Star} \shortauthors{Jeon et al.}

\begin{document}

\title{A New SX Phe Star in the Globular Cluster M15 }

\author{Young-Beom Jeon\altaffilmark{1}, Seung-Lee Kim}
\affil{Korea Astronomy Observatory, Taejeon, 305-348, KOREA \\
Email: ybjeon@boao.re.kr and slkim@kao.re.kr}

\author{ Ho Lee}
\affil{Dep. of Earth Science Education, Korea National University of Education, Choongbuk, 363-791, Korea\\
Email: leeho119@boao.re.kr}

\and

\author{Myung Gyoon Lee}
\affil{Astronomy Program, SEES, Seoul National University, Seoul,
151-742, KOREA
\\ Email: mglee@astrog.snu.ac.kr}

\altaffiltext{1}{Also at the Astronomy Program, SEES, Seoul
National University, 151-742, KOREA }

\begin{abstract}

A new SX Phe star (labelled SXP1) found from $BV$ CCD photometry is the first to be discovered in
the globular cluster M15. It is a blue
straggler and is located $102\arcsec.8$ north and $285\arcsec.6$
west of the center of M15 \citep{har96}.
Mean magnitudes of SXP1 are $<$$B$$>$ = 18$\fm$671 and
$<$$V$$>$ = 18$\fm$445. The amplitude of variability of   SXP1
is measured to be $\Delta V \approx 0.15$. From
multiple-frequency analysis based on the Fourier decomposition
method, we detect two very closely separated pulsating
frequencies: the primary frequency at $f_1=24.630$ c/d for both
$B$- and $V$-bands, and the secondary frequency at
$f_2=24.338$ c/d for the $B$-band and 24.343 c/d for the $V$-band. This
star is the second among  known SX Phe stars found to pulsate
with very closely separated frequencies ($f_2/f_1\ge0.95$).
These frequencies may be explained by excitation of nonradial
modes; however, we have an incomplete understanding of this phenomenon in the case of  SX
Phe stars with relatively high amplitudes. The metallicity-period
and the variability amplitude-period relations for SXP1 in M15
are found to be consistent with those for  SX Phe stars in
other globular clusters.

\end{abstract}

\keywords{Globular clusters: individual (M15 (NGC 7078)) ---
stars: blue straggler
          --- stars: oscillations --- stars: variable stars }

\section{Introduction}

SX Phoenicis stars are  short-period ($<0^d.1$) pulsating
variable stars. They have low metallicities and high spatial
motions typical of Population II \citep{rod20}.
They are located in a blue straggler region in the HR diagram and
within the lowest section of the classical Cepheid instability strip.
The characteristics of these stars are not yet fully explained by
current stellar theories.

Only a few field SX Phe stars are known at present, and
most of the known SX Phe stars were discovered in  galactic
globular clusters and in two dwarf spheroidal galaxies, Carina and
Sagittarius. Recently,  \citet{rod20} published a
catalogue of SX Phe stars in galactic globular clusters including
those in the two dwarf spheroidal galaxies. They listed a total
of 122 SX Phe stars belong to 18 globular clusters and 27
belonging to 2 galaxies,
covering  information published until January, 2000. Since the
initial discovery of SX Phe stars in the globular cluster $\omega$ Cen
\citep{nis81}, the identification rate of these stars in  globular
clusters has increased rapidly in the last decade (see Figure 1
in \citet{rod20}).

In this paper we report the first discovery of an SX Phe star
(hereafter referred to as SXP1) in the globular cluster M15
(RA\,=\,21$^h$ 29$^m$ 58$\fs$3, DEC\,=\,+12$\arcdeg$ 10$\arcmin$
01$\arcsec$, J2000; \citet{har96}). M15 has an extremely low
metallicity [Fe/H] $=-2.25$, an interstellar reddening
$E(B-V)=0.10$, and a distance modulus $(m-M)_V=15.37$
\citep{har96}. There are 126 known variable stars in M15 but no
SX Phe star had yet been discovered in this cluster
\citep{cle20}. Preliminary results from this study were presented
by \citet{jeo20}.

\section{Observations and Data reduction}

\subsection{Observations}

We have obtained $U\!BV\!I$ CCD images of M15 on the photometric
night of
 September 13, 1998,
and a series of $BV$ CCD images of M15 on four nights from
August 12th to 16th, 1999 and over 2 nights from September 25th to
26th, 2000.
A total of 194 (over $\sim$29.7 hours), and 232 (over
$\sim$32.1 hours) frames were obtained for $B$- and $V$-bands,
respectively.
The observation log is listed in Table 1. \placetable{tbl-1}

The CCD images were obtained with a thinned SITe 2k CCD ($2048
\times 2048$ pixels) camera attached to the 1.8m telescope at the
Bohyunsan Optical Astronomy Observatory\,(BOAO). The size of the
field of view of a CCD image is $11\farcm6\times11\farcm6$ at the f/8
Cassegrain focus of the telescope. The readout noise and gain of
the CCD are 7.0$e^-$ and 1.8e$^-/$ADU, respectively. We used the
$2\times2$ binning mode, resulting in a pixel scale of $0.6876$
arcsec per pixel.
A greyscale map of a $V$ CCD image of M15 is shown in Figure 1.

\placefigure{fig1}

\subsection{Data Reduction}

Using the IRAF/CCDRED package, we processed the CCD images to correct
overscan regions, trim unreliable subsections, subtract bias
frames and correct flat field images. Instrumental magnitudes
were obtained using the point spread function fitting photometry
routine in the IRAF/DAOPHOT package \citep{mas92}.

The instrumental magnitudes of the stars in M15 observed on
September 13, 1998 were transformed to the standard system using
photometry of the Landolt standard stars obtained on the same night
as M15 \citep{lan92}. Then the time-series $BV$ data were
calibrated using these data.
Detailed analysis and results of the $U\!BV\!I$ photometry of M15
will be presented elsewhere \citep{jeo21}.

We applied the ensemble normalization technique
\citep{gil88,kim99} to normalize instrumental magnitudes between
time-series CCD frames. We used about a hundred  normalizing
stars ranging from 14\fm0 to 17\fm5 for the $V$-band and from 13\fm5
to 17\fm3 for the $B$-band except for variable stars and central
stars within r $<$ 1\farcm5. The normalization equation is
\begin{equation}
  B ~ or ~ V = m + c_1 + c_2(B-V) + c_3P_x + c_4P_y
\end{equation}
where $B, V$, and $m$ are the standard  and instrumental
magnitudes of the normalizing stars, respectively. $c_1$ is the zero
point and $c_2$ is the color coefficient. $c_3$ and $c_4$ are
used to correct position dependent terms such as  atmospheric
differential extinction and  variable PSF.

\section{Light Curves of the First SX Phe Star in M15 and Frequency Analysis}

After photometric reduction  of the time-series frames,
we inspected luminosity variations for about 21,000 stars
to search for variable stars.
We  confirmed 86 previously known RR Lyrae stars and one
Cepheid variable, and  discovered 16 new variable stars in
the cluster: two faint eclipsing binaries,  two long-period
variable stars, three RR Lyrae stars, eight variable candidates,
and one SX Phe star (SXP1). Here we report only the results on
the SX Phe star, and detailed results on the other variable stars
will be published elsewhere \citep{jeo22}.

There had been hitherto no previously known eclipsing binaries or SX
Phe stars in M15 \citep{rod20,cle20}.  SXP1 is the first SX Phe
star discovered in M15.
SXP1 is located 102\farcs8 north and 285\farcs6 west of the
center of M15, as marked by the V in Figure 1. The coordinates of
SXP1 are RA(2000)\,=\,21$^h$ 29$^m$ 39$\fs$4 and
DEC(2000)\,=\,$+12\arcdeg$ 11$\arcmin$ 43$\farcs$4.

\placefigure{fig2}

$BV$ light curves (dots) of  SXP1 we obtained are displayed in
Figure 2.
The curves are sinusoidal with short periods and low amplitudes
showing that  SXP1 is a pulsating variable star. The maximum
amplitudes of  SXP1 in the $B$- and $V$-bands are estimated to be
$0\fm17$ and $0\fm15$, respectively.
It should be noted that there are amplitude-modulating features in
the light curves of  SXP1, implying the excitation of
closely[ ]separated pulsating frequencies.

We have performed multiple-frequency analysis to find the pulsating
frequencies of  SXP1 using the discrete Fourier transform
(DFT) method and  linear least-squares fitting method
\citep{kim96}. Figure 3 displays the power spectra of  SXP1
for the $B$- and $V$-bands. \placefigure{fig3} The top panel in Figure
3 shows the spectral window, and the other panels represent the
pre-whitening processes.
A primary frequency at $f_1=24.630$ c/d for both bands is
evident. After the primary frequency $f_1$ is prewhitened
(the third panels from the top in Figure 3), the secondary
frequency is detected at $f_2=24.338$ c/d for the $B$-band and 24.343
c/d  for the $V$-band. Since the amplitude signal-to-noise ratios (S/N)
are larger than 4 \citep{bre93}, the secondary frequencies can be
accepted as intrinsic frequencies. After removing synthetic
curves with the two frequencies from the data, the residual light
curves indicate that there are no more frequencies detectable in
the data (see  the bottom panels in Figure 3). The results of the
multiple-frequency analysis for  SXP1 are summarized  in Table
2. Synthetic light curves obtained from this analysis are
superimposed on the data in Figure 2, and show good agreement.
\placetable{tbl-2}

\section{Discussion}

\subsection{SXP1: an SX Phe star or a $\delta$ Sct star?}
In Figure 4, we show the position of  SXP1 in the
color-magnitude diagram (hereafter CMD) of M15.  SXP1 is found
to be located in the blue straggler region
 along an extension of the main sequence, in a region brighter and bluer than
the main sequence turnoff point.
\placefigure{fig4}
The mean magnitudes of  SXP1 are $<$$B$$>$ = $18\fm671$ and
$<$$V$$>$ = $18\fm445$.
Based on the position of  SXP1 in the CMD in conjunction with its
pulsation period and amplitude, it could either be an SX Phe star in
the globular cluster or a field $\delta$ Scuti star. In order to
define the pulsating type of  SXP1 more clearly, we examined the
$V$-amplitude versus period diagram for SX Phe stars and $\delta$
Scuti stars in Figure 5.
\placefigure{fig5}
The sources of the data
are \citet{rol20} for field SX Phe stars and  $\delta$ Scuti
stars, and \citet{rod20} for SX Phe stars in  galactic
globular clusters. Figure 5 shows that the $V$-amplitude and
period of  SXP1 is consistent with those for other SX Phe
stars in  globular clusters and that the $V$-amplitude of
SXP1 is much larger than those of  $\delta$ Scuti stars with
the same period. This shows that  SXP1 is an SX Phe star, not
a $\delta$ Scuti star.

\subsection{Membership of  SXP1}

Following the suggestion of \citet{mcn97} that SX Phe stars with
$\Delta$$V$ $\le 0\fm20$ can be classified as  first-overtone
pulsators, we can assume that  SXP1 is a first-overtone
pulsator. After fundamentalizing the dominant frequency $f_1$ by
assuming a fundamental to first overtone period ratio
$P_1/P_0=0.778$, we obtain the absolute magnitude of  SXP1,
$M_V=2\fm84$ and the distance modulus, $(m-M)_V=15\fm$61 using
the period-luminosity ($P-L$) relation given by McNamara (1997;
his Eq.(4), $M_V = -3.725 \log P_0 - 1.930$ ).

Recently, \citet{mcn01} established  the equations for the magnitudes
of the horizontal branch and the main-sequence turnoff:
$M_V(HB) = 0.30[Fe/H] + 0.92$ and $M_V(TO) = 0.34[Fe/H] - 4.48$.
Using these equations we obtain $M_V(HB)$ = 0.24 and $M_V(TO)$ = 3.71
for [Fe/H] = --2.25, the metallicity of M15.
From the color magnitude diagram of M15 (see Fig. 4),
the magnitudes of the horizontal branch and the main-sequence turnoff are,
respectively, $V(HB)$ = 15.80 and $V(TO)$ = 19.30.
Therefore, the corresponding absolute magnitudes of
SXP1 are $M_V$ = 2\fm88 from $M_V(HB)$ and 2\fm85 from $M_V(TO)$.
These are in good agreement with the absolute magnitude of SXP1, $M_V$ = 2.84
derived from the P-L relation.

If we use the
period-luminosity-metallicity ($P - L - $ [Fe/H]) relation given
by \citet{nem94}, $M_V=-2.56 \log P_0 + 0.32$[Fe/H] $ + 0.36$ ,
we obtain the distance modulus of $(m-M)_V = 15\fm$49, adopting
the cluster metallicity [Fe/H] $=-2.25$ \citep{har96}. These two
results are consistent within 2$\sigma$ (for $P-L$ relation) or
1$\sigma$ (for $P - L - $ [Fe/H] relation) error with the
distance modulus of M15, $(m-M)_V=15\fm$37 $\pm0\fm15$
\citep{har96}.
The fundamentalized $f_1$ period and [Fe/H] relation of  SXP1
is also consistent with the metallicity-period relation of other
SX Phe stars in the galactic globular clusters \citep{rod20},
as shown in Figure 6.
All of these facts support the contention that  SXP1 is a member of M15, and
also an SX Phe star in the cluster.

\subsection{Two close frequencies of  SXP1}

It should be noted that the two detected frequencies of  SXP1 are
very closely separated
 (frequency ratio = 0.988).
This is often seen in the case of low-amplitude $\delta$ Scuti
stars, but it is very rare for SX Phe stars. Up to now, only one
SX Phe star, BL Cam \citep{zho99}, among the known SX Phe stars
is known to have very closely separated frequencies (frequency
ratio $\ge$ 0.95).
These frequencies can be explained by excitation of a nonradial
mode \citep{zho99}. However, the excitation of nonradial modes has
not yet been physically understood in the case of SX Phe stars
with relatively high amplitudes. Recently, nonradial pulsation
components were also detected from the frequency analysis of
first overtone RR Lyrae stars
\citep{alc20}. To identify the pulsation mode of  SXP1, we
tried to obtain the phase differences between the color index
($B-V$) and the $V$ magnitude \citep{gar20}, but failed because
our data are not of sufficient quality to obtain the variation of color index.
Better data are needed to identify the pulsation mode of  SXP1.

\placetable{tbl-3}

\section{Summary}

We have discovered the first SX Phe star (SXP1)
 in the globular cluster M15 from $BV$ CCD photometry.
Table 3 summarizes physical parameters of  SXP1 derived  in
this study. Two very closely separated frequencies are detected
in the light curves of SXP1, which could be  explained by
excitation of a nonradial mode.

\acknowledgments

We are grateful to the referee, D. H. McNamara for useful comments.

\clearpage

\plotone{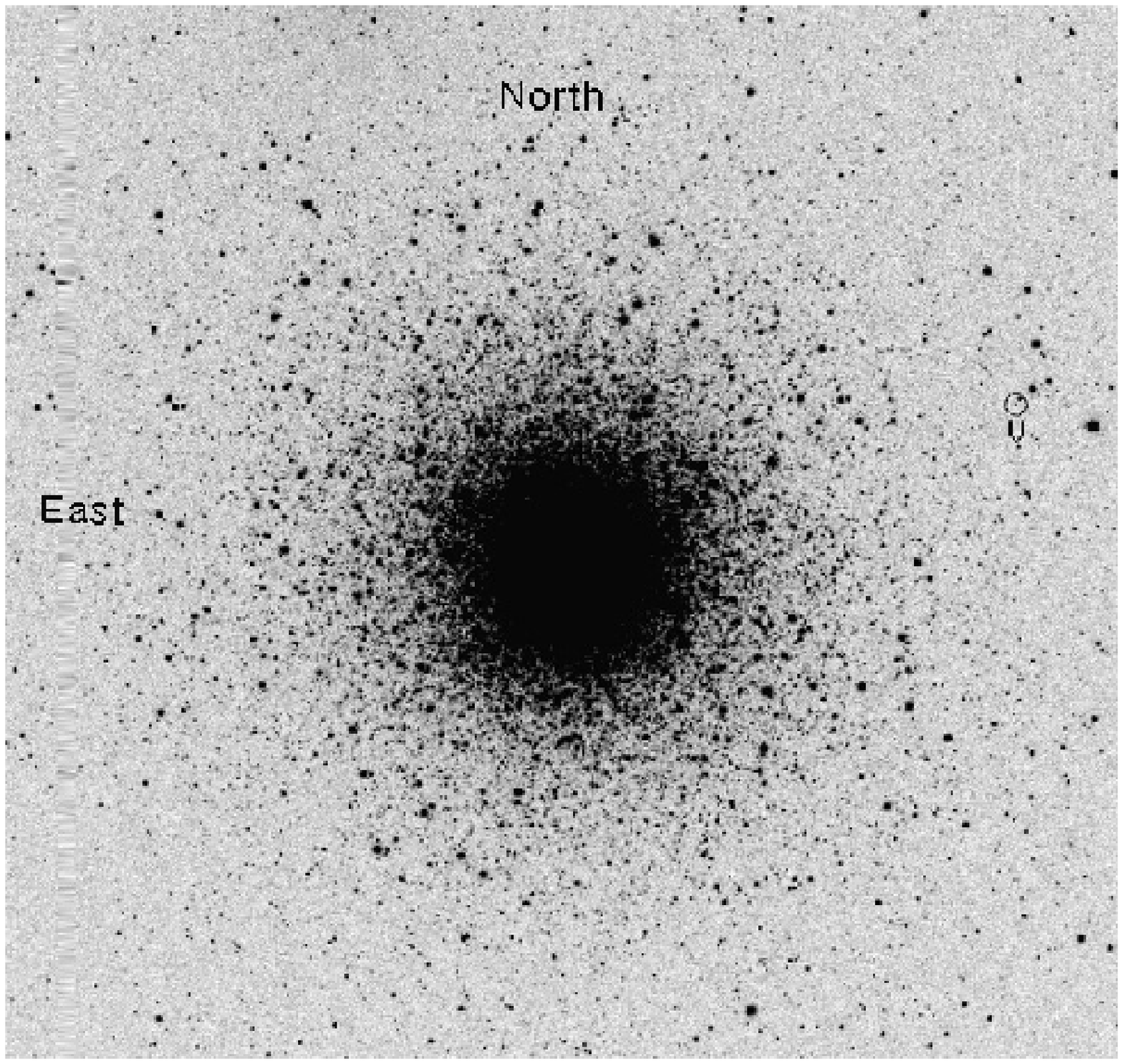}
\figcaption[Jeon.fig01.ps]{ A greyscale map of a
$V$-band CCD image of the globular cluster M15. A new SX Phe star
(SXP1) is denoted by V, in the center of a small circle. SXP1
is located 102\farcs8 north and 285\farcs6 west of the center
of M15. \label{fig1}}

\plotone{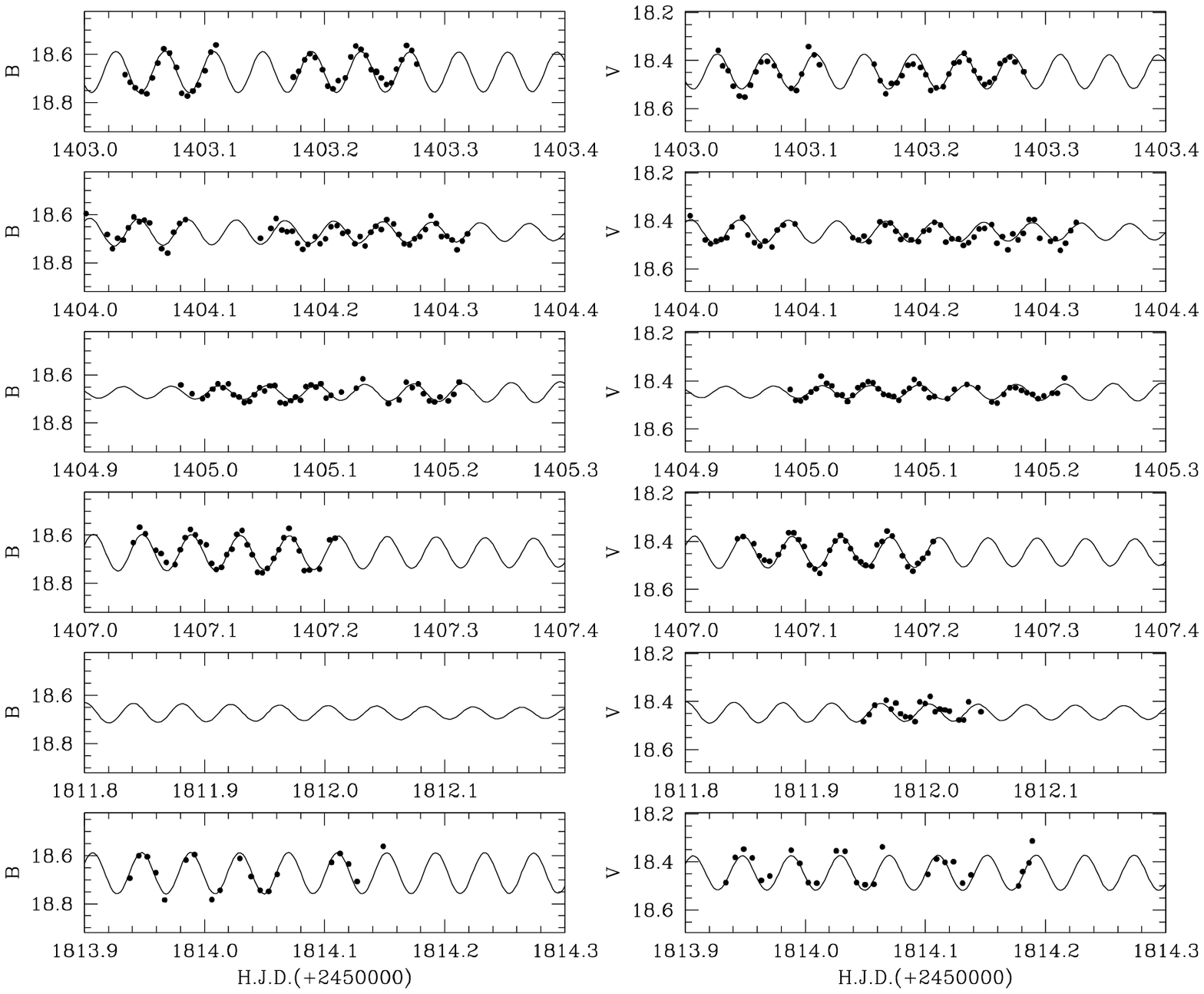}
\figcaption[Jeon.fig02.ps]{ Observed light curves
(dots) for SXP1 for B-band (left) and V-band (right).
Synthetic light curves (solid lines) obtained from the
multiple-frequency analysis (see Table 2) are superimposed on the
data. \label{fig2}}

\plotone{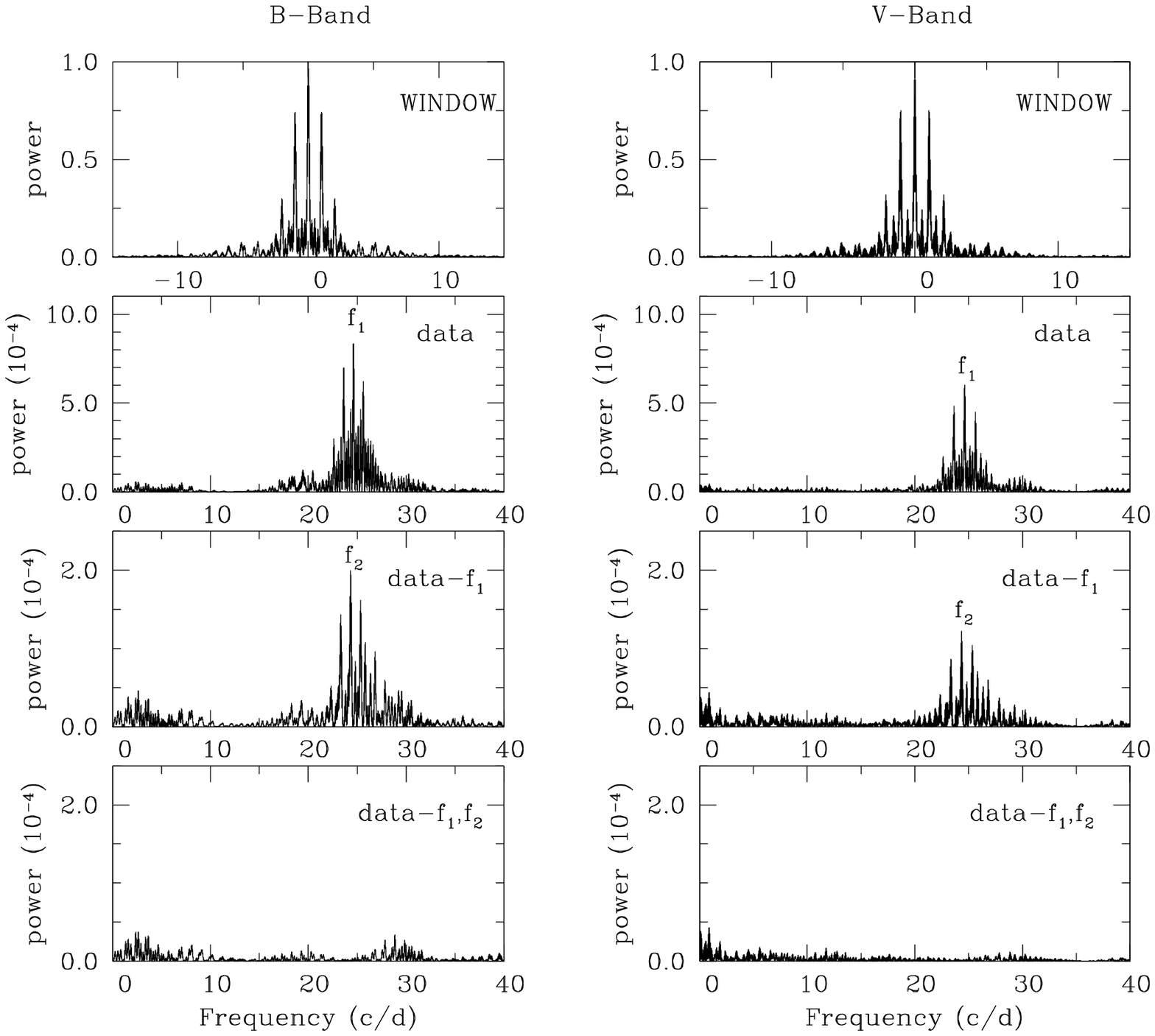}
\figcaption[Jeon.fig03.ps]{ Power spectra of
SXP1 for B-band (left) and V-band (right). Window spectra are in
the top panel. Two closely-separated frequencies, $f_1$ and
$f_2$, are clearly found. \label{fig3}}

\plotone{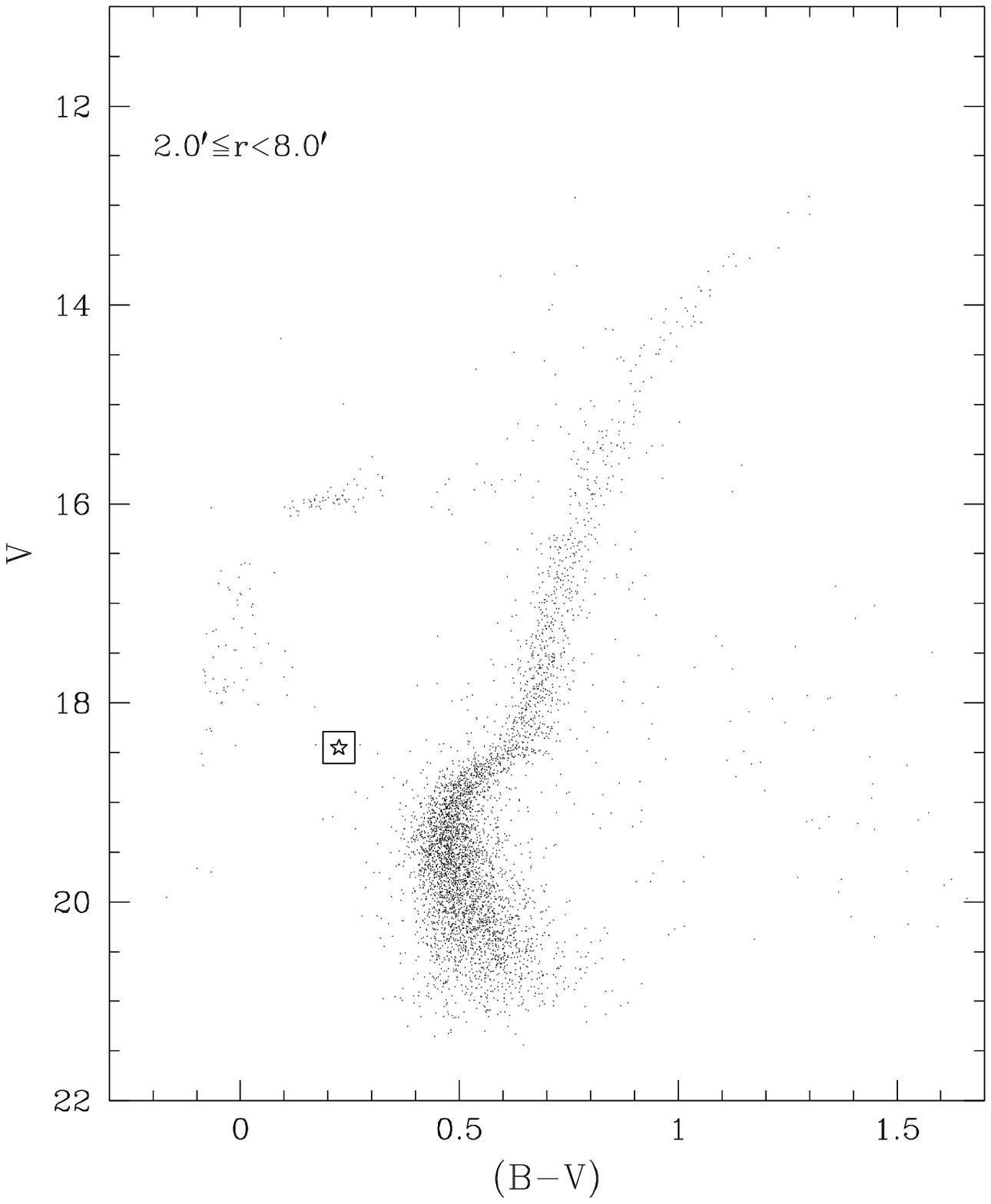}
\figcaption[Jeon.fig04.ps]{ Position of  SXP1
in the color-magnitude diagram of M15. Note that it is located in
the blue straggler region. \label{fig4}}

\plotone{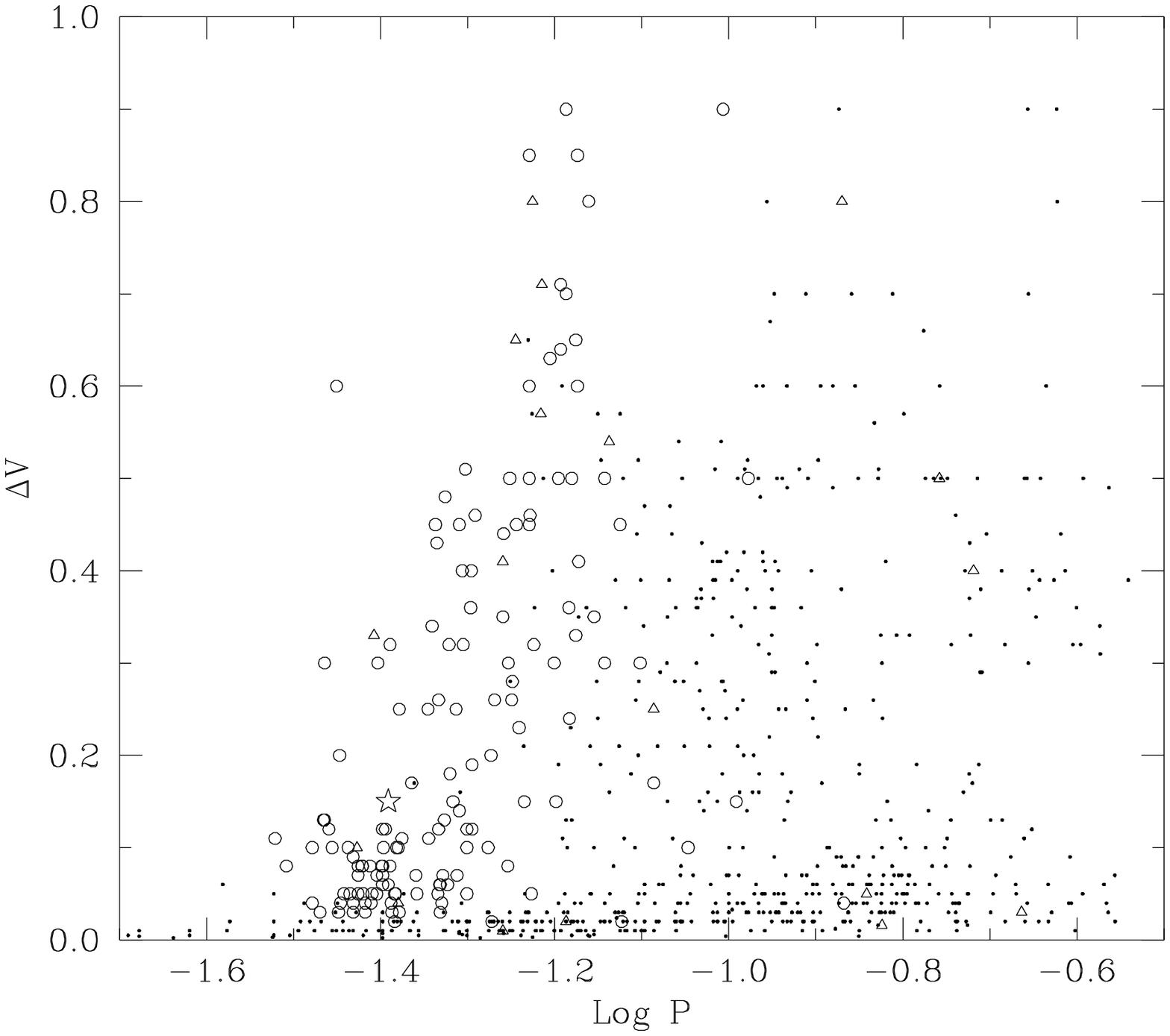}
\figcaption[Jeon.fig05.ps]{ V amplitude versus
period diagram\,;\, a star symbol denotes] SXP1 in M15,
small dots denote $\delta$ Scuti stars,
open triangles respresent field SX Phe stars,
and open circles indicate SX Phe stars in other globular clusters.
\label{fig5}}

\plotone{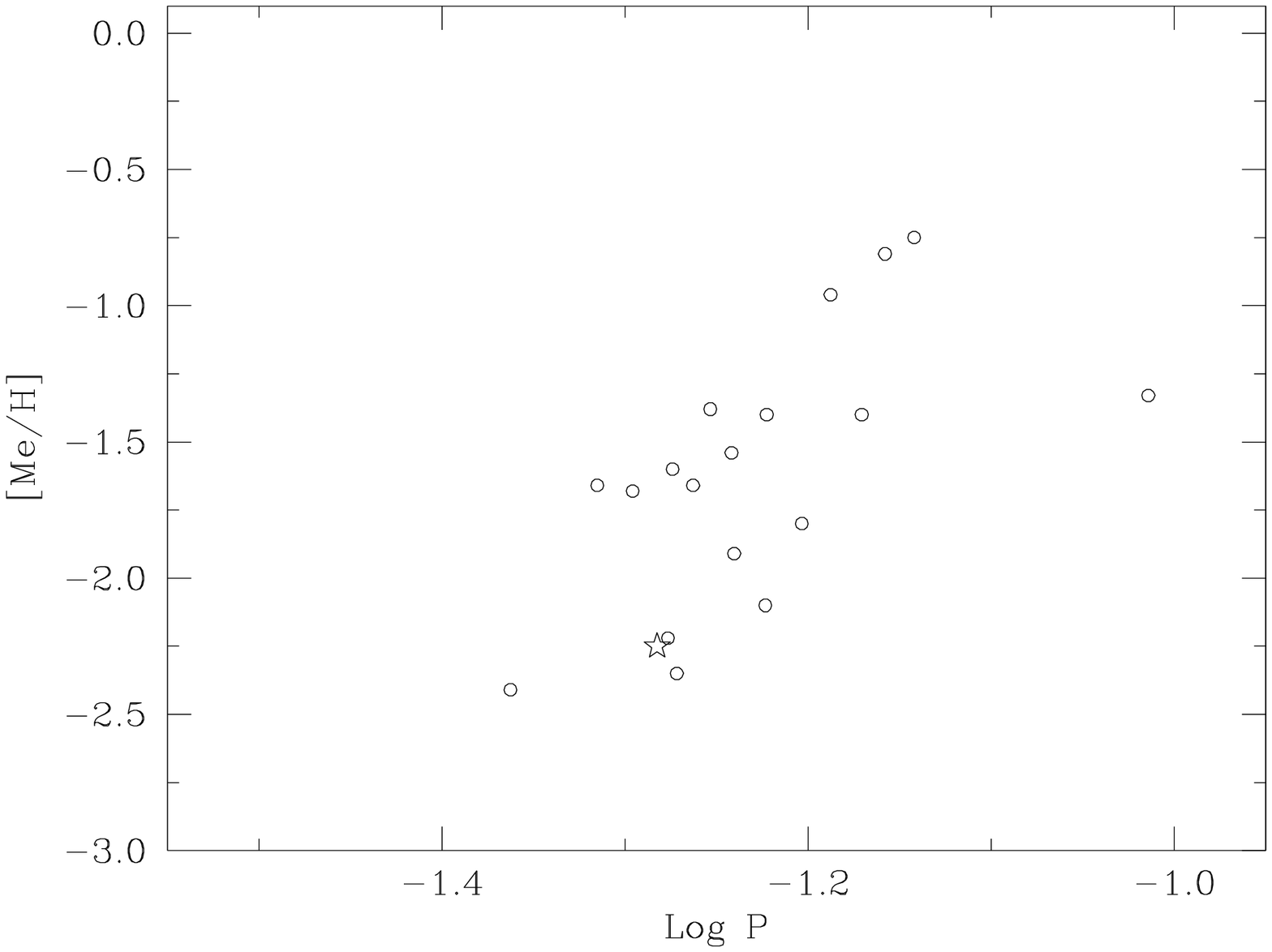}
\figcaption[Jeon.fig06.ps] {[Me/H] versus
fundamentalized period diagram for SX Phe stars in galactic
globular clusters. A star symbol represnts SXP1. \label{fig6} }
\clearpage

\begin{deluxetable}{rlrclcll}
\tablecaption{Observation log. \label{tbl-1}} \tablewidth{0pt}
\tablecolumns{8} \tablehead{ \colhead{Date}&\colhead{}    &
\colhead{Start H.J.D.}& \colhead{Duration} &
\colhead{N$_{obs}$(Filter)} & \colhead{Seeing}  & \colhead{Exposure Time}     & \colhead{Remark}\\
&&&  & & & } \startdata
Aug. 12, & 1999 &        2451403.03 & 6\fh0 & 46$(V)$, 43$(B)$ & 2\farcs4 & V100s, B200s & nonphotometric \\
     13, &      &            404.00 & 7\fh9 & 56$(V)$, 53$(B)$ & 2\farcs2 & V100s, B200s & nonphotometric \\
     14, &      &            404.99 & 5\fh5 & 47$(V)$, 43$(B)$ & 2\farcs0 & V100s, B200s & nonphotometric \\
     16, &      &            407.04 & 4\fh1 & 36$(V)$, 36$(B)$ & 2\farcs2 & V100s, B200s & nonphotometric \\
Sep. 24, & 2000 &            811.95 & 2\fh4 & 21$(V)$          & 2\farcs8 & V200s        & cirrus \\
     26, &      &            813.93 & 6\fh2 & 26$(V)$, 19$(B)$ & 2\farcs2 & V200s, B360s & nonphotometric \\

\enddata

\end{deluxetable}

\clearpage

\begin{deluxetable}{crrrrrrrrr}
\tablecaption{Results of the multiple frequency analysis.
\label{tbl-2}} \tablewidth{0pt} \tablecolumns{10} \tablehead{
\colhead{} & \multicolumn{4}{c}{$B-$band} & \colhead{} & \multicolumn{4}{c}{$V-$band}  \\
\cline{2-5}\cline{7-10} \colhead{} & \colhead{Freq.(c/d)} &
\colhead{Amp.\tablenotemark{a}}   &
\colhead{Phase\tablenotemark{a}}  &
\colhead{S/N\tablenotemark{b}}  & \colhead{} &
\colhead{Freq.(c/d)} & \colhead{Amp.\tablenotemark{a}}   &
\colhead{Phase\tablenotemark{a}}  & \colhead{S/N\tablenotemark{b}}  \\
} \startdata
$f_1$ & 24.630  & 0\fm054  &$-$0.240 &  9.2 & & 24.630 & 0\fm048 & $-$0.324 & 10.4\\
$f_2$ & 24.338  & 0\fm032  &   4.117 &  7.2 & & 24.343 & 0\fm025 &    3.919 & 6.9 \\
\cline{1-5}\cline{7-10}
s.d.\tablenotemark{c} & \multicolumn{4}{c}{0\fm022} &  & \multicolumn{4}{c}{0\fm022}  \\
\enddata

\tablenotetext{a}{$B$ or $V = Const + \Sigma_j A_j \cos \{2 \pi
f_j (t - t_0) +
     \phi_j\},~~~ t_0 = H.J.D. 2451400.00$. }
\tablenotetext{b}{Amplitude signal-to-noise ratio introduced by
Breger et al. (1993). } \tablenotetext{c}{Standard deviation
after fitting synthetic curves to the data. }

\end{deluxetable}

\clearpage

\begin{deluxetable}{ccccccc}
\tablecaption{Physical properties of the SXP1. \label{tbl-3}}
\tablewidth{0pt} \tablecolumns{8} \tablehead{
\colhead{RA(2000)}&\colhead{ DEC(2000)} & \colhead{$<$$V$$>$}&
\colhead{$<$$B$$>$ -- $<$$V$$>$} &
\colhead{Period} & \colhead{$f_2$/$f_1$}   & \colhead{Remark}    \\
} \startdata 21$^h$ 29$^m$ 39$\fs$4 & $+12\arcdeg$ 11$\arcmin$
43$\farcs$4 & $18\fm445$ &  $0\fm226$ &
0$\fd$0406 & 0.988  & nonradial oscillator \\
\enddata

\end{deluxetable}

\clearpage

\end{document}